\documentclass[sigconf]{acmart}
\AtBeginDocument{%
  }

\copyrightyear{2026}
\acmYear{2026}
\setcopyright{cc}
\setcctype{by}
\acmConference[EASE '26]{International Conference on Evaluation and Assessment in Software Engineering}{June 09--12, 2026}{Glasgow, United Kingdom}
\acmBooktitle{International Conference on Evaluation and Assessment in Software Engineering (EASE '26), June 09--12, 2026, Glasgow, United Kingdom}
\acmDOI{10.1145/3816483.3816621}
\acmISBN{979-8-4007-2348-3/2026/06}

\begin{document}

\title{Evaluating LLM-Generated Code: A Benchmark and Developer Study}


\author{Joanna Szych}
\affiliation{%
  \institution{IU Internationale Hochschule Bad Honnef}
  \city{Gdańsk}
  \country{Poland}}
\email{joanna.szych@iu-study.org}
\orcid{0009-0008-1154-9046}

\author{Anne Schwerk}
\authornote{Corresponding author}
\affiliation{%
  \institution{IU Internationale Hochschule Bad Honnef}
  \city{}
  \country{Germany}}
\email{anne.schwerk@iu.org}
\orcid{0000-0002-5014-5289}

\renewcommand{\shortauthors}{Szych \& Schwerk}

\begin{abstract}
Code generation is one of the tasks for which the use of Large Language Models is widely adopted and highly successful. Given this popularity, there are many benchmarks dedicated to code generation that can help select the best model. However, they primarily focus on measuring solution correctness, leaving other aspects, such as code quality and usability, behind. This paper aims to describe a custom tree-fold evaluation methodology for code generated by Large Language Models that bridges this gap. The methodology includes a dedicated correctness benchmark based on a complex multi-level computer science project, code quality verification, and a survey of developers’ opinions on generated code samples gathered through a structured code-review process. The proposed methodology's usage and usefulness are demonstrated by evaluating and comparing three general-purpose Large Language Models: GPT-4.1, DeepSeek-V3-0324, and Claude Opus 4. The results show that reviews gathered from developers can yield many new findings, especially those related to the code being in a production-ready state, that would not be possible to obtain using the standard correctness-focused benchmark approach.  
\end{abstract}



\maketitle

\section{Introduction}

In the code generation task (NL2Code), the most common evaluation method is a benchmark that measures the percentage of predefined tasks solved correctly, verified by unit tests. A characteristic of benchmarks is their structure, where each task focuses on a single assignment. Thanks to this, tasks are independent of one another and can be executed in any order. However, real-world coding assignments require a very different approach to programming. In many personal and work-based use cases, it is often necessary to create a slightly larger codebase than a simple script to solve the chosen task. In such cases, when one wants to use a Large Language Model (LLM) for help, it may be necessary to ask for different functionalities in separate prompts. That means that solving some projects with LLMs, especially from the beginning, may require a longer conversation with the model. Unfortunately, most benchmarks do not assume this type of sequential prompting for the creation of whole projects, especially larger ones with many files and dependencies. Moreover, when working on such projects, it is common for an author to share the code with others or pick it back up after a few months' break. This is why, in real-world coding projects, in addition to working correctly, the code should be easy to read, understand, and maintain. While some common good practices can be verified using static analysis, many aspects remain difficult to detect without a human eye. That is one of the reasons why, in many professional scenarios, a newly created code is verified by at least one other person, who checks whether it follows agreed coding standards. Unfortunately, there is a scarcity of research in this area that involves human experts evaluating code generated by Large Language Models.

This paper aims to introduce an evaluation methodology that bridges the aforementioned research gaps by developing a reproducible approach to using LLMs for complex project creation and verifying the quality of those solutions through human evaluation. To reach this goal, the following three research questions were posed as a means of guiding and summarising further work: 
\begin{description}
\item[\texttt{RQ1}:] How can we design an evaluation framework that assesses LLM-generated code in realistic, multi-component project contexts rather than isolated function-level tasks?
\item[\texttt{RQ2}:] How can we combine automated correctness testing and human quality assessment to comprehensively evaluate LLM-generated project implementations?
\item[\texttt{RQ3}:] To what extent do automated correctness metrics and human quality assessments align and complement each other when evaluating LLM-generated projects?
\end{description}
To address the research questions, a custom three-fold evaluation methodology was developed that combines the complexity of the generated code with human feedback on its quality. It is based on a multi-part computer science project, which LLMs are asked to solve through a series of predefined prompts. To verify the correctness of the prepared solution, a small custom benchmark was created for this assignment. The results were combined with a code quality score from dedicated software for a comprehensive overview of the solution. In the last step, the generated code undergoes a structured code review conducted by volunteer developers using a specially designed questionnaire. Presenting this approach leads this paper to make the following contributions to the community:
\begin{itemize}
\item A three-fold evaluation methodology framework combining automated testing with code quality verification and developer surveys to capture both correctness and usability
\item A project-level correctness benchmark to assess generated software implementations
\item A survey design dedicated to gathering code reviews from developers in a structured manner 
\item A demonstration of the framework through the evaluation of three LLMs on a Python project
\end{itemize}
The details of the listed contributions are presented in the following sections. First, an overview of related work is presented to provide background on what has already been discovered in the field of code generation evaluation. After that, the creation and setup of each element contributing to the designed evaluation methodology are described. This is followed by an overview of experimental results based on three general-purpose LLMs: GPT-4.1, DeepSeek-V3-0324, and Claude Opus 4. Finally, the work's limitations and lessons learned from the experiments are discussed, followed by a conclusion.

\section{Related Work}
To provide an objective and meaningful literature review, which sets a solid background for future deliberations, guidelines defined in “Preferred Reporting Items for Systematic Reviews and Meta- Analyse" (PRISMA) methodology were applied. The search strategy was to select articles written in English from six reputable databases, queried using the keywords “LLM code generation”, “LLM comparison methods”,  “LLM evaluation benchmarks”, and “LLM code generation benchmark”, combined with the “OR” operator. The articles were either included or excluded from the subsequent stage of the process based on whether they revolved around comparing models' coding capabilities in standard programming languages or not. From each source, the first 15 results from the years 2018-2025 were chosen for further examination, which gives 90 results. From those, 73 were removed due to duplicated titles or being outside the scope, leaving only 17 relevant articles. A few of those selected papers are surveys, which summarise all discoveries in the area of Large Language Models for the code generation task from recent years. This type of literature contains an extensive list of resources, from which an additional 26 papers were categorised as containing extremely useful information about popular code generation benchmark and evaluation approaches. The following section synthesises the knowledge gathered in 43 selected papers and highlights the most relevant findings.

\subsection{Correctness evaluation}
\label{sec:subsection}
One of the basic dimensions of evaluating code generation, and what most of the currently available benchmarks aim to measure, is the correctness of the generated solution. Some of the most popular benchmarks, such as HumanEval \cite{humaneval}, HumanEval+ \cite{humanevalplus}, or MBPP \cite{austin2021MBPP}, are often based on Python code generation. However, there are also expanded versions of those datasets, such as Multilingual HumanEval \cite{MBXP}, HumanEval-X \cite{HumanEval-X}, or MBXP \cite{MBXP} that include tasks in other programming languages. When running one of such benchmarks, models are queried to perform all tasks defined in it, and afterwards, each generated solution is evaluated using a set of specially prepared tests. The percentage of passed cases is treated as a benchmark score. Most of the available benchmarks focus on unit-level code generation, which includes problems that can be solved by implementing or completing a single programming function, or, more generally, a short, few-line-long piece of code. Such an approach, while necessary to establish a level of function-complexity that new LLMs can solve, is far from a real-world programming scenario in which those LLMs could be used. To bridge this gap, some researchers have presented higher-level benchmarks, focusing on class-level evaluation \cite{ClassEval} or on capabilities for solving real GitHub bugs \cite{CoderEval}, \cite{CONCODE}. Another level of complexity was reached in benchmarks for LLMs operating in a repository-wide context, such as RepoBench \cite{repobench}, SWE-bench \cite{swebench}, or RepoEval \cite{repoeval}. Nevertheless, none of those approaches addresses the very important programming task that developers may encounter in their work: building a new project from the ground up. There is a lack of data on how effectively the models can be used to build the solution when relying only on prompts that describe the requirements, without any starting code or prerequisites for specific functions, methods, or classes. Among all selected papers, only one work addressed this idea, which is a study done by McDanel \& Novak \cite{NIFTY_study}, who used the SIGCSE Nifty Assignments list, a collection of computer science problems designed for student use, as a benchmark to identify areas where Large Language Models lack expertise. As shown by experiments across three LLMs and six assignments of varying complexity, LLMs struggle with problems involving visual elements or requiring DevOps competencies. This work demonstrates that such research, based on full-scope assignments, can offer an interesting view of the capabilities of Large Language Models in a realistic setting, as many programmers know.

\subsection{Quality evaluation}
\label{sec:subsection}
When working on a project, especially one that may be further shared or evolved over a longer period of time, the generated code must not only be correct but also easy to understand, modify, and reuse. Those dimensions are often a part of code quality evaluation. While such an evaluation is often connected with automatic verification based on a set of rules, a common way to verify if a given piece of code is easily readable and understandable is a human-based code review. The process places a human element at the centre and includes gathering feedback from experienced developers. 
Unfortunately, this code-verification practice is not applied in research evaluating LLM-generated code, as only a limited number of studies incorporate human factors into such evaluations. One of the works that applies this idea is the User-Centric Evaluation study by Miah \& Zhu \cite{UserCentricEval}, whose authors have asked one of the master's students in data analytics to rate each LLM-generated solution on a scale of 1 to 5 in categories such as logic clarity, readability, and structuredness. A separate study by Wang et al. \cite{WWang2024} involved 109 human participants and focused on comparing developers' performance with and without ChatGPT assistance while solving two coding tasks. Both studies introduce many important findings for the field of evaluating the usefulness of LLMs for human users, such as a positive attitude among participants towards the use of ChatGPT and its potential to replace code-searching practices, especially for simple puzzles. Unfortunately, conducting such a series of experiments is generally time-consuming and requires significant effort, which may be a reason why evaluation methods involving human participants are rarely reported. Nevertheless, the need to evaluate the quality of LLM-generated code remains, and some works attempt to fill this gap by using the LLM-as-a-Judge methodology \cite{judgingllmasajudge}. One example of that is the CompassJudger-1 model \cite{farchi2024}, an LLM specifically retrained to perform tasks such as comparing the answers of two other models or generating an analysis of those answers. While this approach is based on comparison, another approach uses LLM-as-a-Judge to evaluate the usefulness of code samples generated by models tasked with performing the NL2Code task. Even though using models to evaluate outputs generated by other models is yielding promising results and can reduce the time and costs of human evaluation, the question of “Can LLMs Replace Human Evaluators?” remains valid. A study with exactly this title was recently presented by R. Wang et al. \cite{Can_LLMs_Replace_Human_Evaluators}, who verified this assumption in the area of Software Engineering. The paper explores seven different methods for using the LLM-as-a-Judge approach and finds that output-based comparison has the highest correlation with scores manually assigned by human judges, which leads to the conclusion that “LLM-as-a-judge methods can potentially replace human evaluations in certain SE tasks" \cite{Can_LLMs_Replace_Human_Evaluators}. However, it is important to note that the selected LLMs exhibit highly variable accuracy across tasks, and the scores were compared to those assigned by only two human evaluators. While yielding highly important findings, this work does not provide proof that the LLM-as-a-Judge methodology can be used safely in place of human factors in the evaluation process.

\subsection{Summary and research gaps}
\label{sec:subsection}
A conclusion drawn from the presented literature underscores the importance of continually improving methods for evaluating the outputs of Large Language Models. The area of analysis that includes human experts' verification is uncommon in research due to the time-consuming process of recruiting participants and conducting experiments. Additionally, many available correctness benchmarks focus on tasks with complexity limited to function-level or class-level, and do not assess the ability to solve full-scope projects. This paper aims to address both of these aspects. 

\section{Research methodology}
To answer the three research questions presented in the introduction, a methodology that combines both correctness testing and human-based quality verification is necessary. For the aim of the third research question, both parts must focus on the same generated code to ensure a fair assessment of their alignment. Additionally, as discussed previously, the selected coding assignment should be a relatively complex, multi-component project in which the LLM must remain in context throughout the whole solution process. Given those requirements, there is a lot of room to assess its usability, which may focus not only on code quality itself but also on the project's composition. The details of the chosen assignment and the ways of measuring correctness and quality of the generated solution are discussed in the following subsections.

\subsection{Coding assignment}
The first step in the process is to define the project, which will be solved by a Large Language Model. The selected assignment must reflect the level of complexity expected in real-life scenarios, for example, in developer roles or in the last year of informatics studies. There may be different views on which coding aspects are difficult, but a few areas that consistently make problems more challenging can be identified. One of those areas is the project's scale: assignments that span more topics and have many requirements often require defining many files, classes, and methods to meet those demands, making the project more abstract and harder to navigate. Another area is the uncommonness of the topics and the specific knowledge required to solve them. Many benchmarks and tests focus on relatively common coding topics, such as implementing a function to calculate the Fibonacci sequence, sorting algorithms, and simple API design. Exemplary solutions or tutorials for this type of task are commonly available online and can be included as part of the training data corpus. Even in cases that seem large, such as implementing numerous API endpoints, the necessary knowledge is often easily available. The difficulty usually arises from unknown waters and specialist assignments, when there are limited resources to gather the required information. The last area that may often introduce an additional level of complexity is the destination of written code, as the code that needs to be run in production as a part of a service for other people may require a different setup and solutions than a project that will only be run locally for the needs of one person.

Finding a problem that captures the described characteristics of a challenging task is nontrivial. However, taking the inspiration from the work of McDanel \& Novak \cite{NIFTY_study}, the Nifty Assignment webpage was a place to look for such a problem. The chosen assignment, called “Building the Tree of Life from Scratch,” was created by Dr Christopher Tralie. The project touches on bioinformatics, as it aims to create a phylogenetic tree based solely on the raw DNA sequences of selected species and to capture the evolutionary relationships among them \cite{NIFTY_webpage}. According to the author, this task is usually assigned as the second-to-last project in Data Structures and Algorithms courses and requires substantial knowledge from students. The subjects that are necessary for a correct implementation of the solution include dynamic programming, recursion, graphs, and tree structures. The project is divided into three logical parts: calculating distances between species using the Needleman-Wunsch algorithm, building a phylogenetic tree based on those distances, and performing clustering to identify groups of similar species. This division makes it easy to break it into a few smaller subproblems, allowing the solution to be built gradually and creating logical checkpoints to catch errors early in the development process. However, to complete this assignment, it is also necessary to solve distinguished subproblems in a certain, sequential order. Because of this, the whole project needs to be created in a single session of working with the LLM, mirroring the flow of a developer chatting with the LLM to generate a solution. An additional requirement for this project, added to increase its complexity, was to prepare the program in a production-ready way. This means that the code should be clean, properly commented, and documented in a README file, and, most importantly, all aspects that may require configuration should be easily adjustable. 

The described assignment should be an appropriate proxy for real-world coding tasks, as it includes all of the mentioned challenging aspects. It requires substantial knowledge of a specific topic that is not readily available online; it comprises many smaller tasks, which may require submodules to be solved, thereby increasing the problem's complexity, and it should be prepared in a production-ready manner. Additionally, no starter code or template will be provided to the model, so the goal for the LLM is to produce functional software based solely on the information in the prompts. To cover all parts of the project, a fixed set of 14 prompts was created. They were designed, to the best of the author's knowledge, in a way that a typical developer with an understanding of prompt engineering could write them to solve such an assignment. Each prompt is divided into three parts: the prompt goal, instructions for what the LLM needs to do, and the expected output, with additional information about the output format included. Since it is usually best to give an LLM one major task per request, the prompts were designed accordingly, with some focusing solely on code generation and others adding context to help the LLM better understand the problem.

\subsection{Correctness testing}
\label{sec:subsection}
As previously mentioned, the prompts given to the LLM have no starter code, no function definitions, and no expected project design format. Because of this, the model has complete autonomy over how the solution will be prepared. This means that the only basis for the evaluation process is the program's outcome. However, this also means that the entire verification could have only two outcomes, turning it into a true/false problem. Since the assignment consists of many subproblems, an error in implementing one of them automatically results in the entire solution receiving a negative evaluation. In such a situation, it would be extremely hard to use a score like this to compare models and draw any insights into their capabilities. Due to this limitation, additional points of verification needed to be introduced.

\subsubsection{Benchmark creation} 
\label{sec:subsubsection}
In the original project description on the NIFTY webpage \cite{NIFTY_webpage}, the author of the assignment proposes creating a checkpoint in the program after calculating the Needleman-Wunsch score for pairs of species. The reasoning is that the calculation can take a few minutes, slowing the development of subsequent parts of the solution. To speed up the work, it is recommended to save this part of the solution in a JSON file and read the data from it later. Although created for a different purpose, such a file of intermediate results can also serve as a good checkpoint for correctness testing. Based on this idea, other locations where such checkpoints could be established were identified. Since the first control point was set up after the first bigger subproblem in the whole assignment- the Needleman-Wunsch algorithm - it was natural to look for other ones after another two major steps: the phylogenetic tree creation and clustering task. For the second one, a natural idea was to save the threshold value and species for each cluster in a file, so this requirement was added to the prompts. However, finding an easy-to-compare way to store the intermediate results after the tree-creation task proved somewhat more challenging. One way to save a tree was already presented in the project description: drawing it as a dendrogram and saving it as a PNG file. However, because there is no standard way to automatically assess the quality of the created visualisation, it is difficult to use it to compare different trees. A better option for this task is the Newick format, which is a standard way to represent phylogenetic trees in text. Newick syntax allows the tree structure to be written in many different ways, depending on how much detail one wants to include. For this assignment, two versions of the Newick format are used. The first one records only named leaf nodes, which should be enough to verify the correctness of the node connections. In the second one, both leaves as well as their distances to their direct roots, which should allow to validate proper usage of Needleman-Wunsch scores to create the tree. This requirement was also included in the predefined prompt set. 

Using the described checkpoints in the project outline, it was possible to create a set of seven tests listed below, where each one verifies the correctness of: 
\begin{itemize}
\item Needleman-Wunsch scores
\item Newick syntax with nodes only
\item Newick syntax with nodes and distance
\item Phylogenetic tree structure, based on Newick syntax with nodes only
\item Phylogenetic tree structure, based on Newick syntax with nodes and distance
\item Created clusters based on given thresholds
\item Created PNG file with dendrogram
\end{itemize}
Introduction of those additional checks should allow to look more granularly at the whole solution generated by the model. Although seven test cases may not seem like a large benchmark, they verify the most important aspects of the project, enabling a thorough comparison of the solutions produced by different models.

\subsubsection{Reference solution} 
\label{sec:subsubsection}
The test cases described above are based on verifying the program's output after a specified number of steps. Three of the presented tests, namely the two related to the correctness of Newick syntax and one that checks the existence of a PNG file, can be evaluated by themselves. However, the remaining four tests were designed with a comparison methodology in mind. Because of that, to verify that Needleman-Wunsch scores were properly calculated, ground-truth scores against which other scores will be compared are needed. 

Unfortunately, the NIFTY assignments webpage \cite{NIFTY_webpage} is intended to provide descriptions of assignments, which can be used in computer science studies. This means that sharing an implementation that already solves the project would undermine the webpage's goal. However, the authors are encouraged to provide tools that guide both fellow educators and students in completing the assignments successfully, such as sample data or demos. For the "Building The Tree of Life from Scratch" task, the provided points of verification are: the final visualisation of the dendrogram, the clustering results for two fixed threshold values, and an interactive applet for comparing the similarity scores obtained by the Needleman-Wunsch algorithm. While these are useful and should be sufficient for students completing this task to verify the correctness of various steps during implementation, they are not feasible for large-scale use. As a result, a reference implementation of the solution was required to meet all defined requirements and serve as the basis for creating test cases. 

A program with such a baseline implementation was written by one of the authors of this paper and placed in the dedicated repository \cite{Complex_benchmark}. Since this project will be used to evaluate solutions generated by LLMs, no code generated by any model was used in the making of this reference implementation to avoid including unnecessary mistakes that may be uncaught during testing. The program was written in Python by a single person, with only the support of examples provided by Tralie on the Nifty website \cite{NIFTY_webpage}. The completed application works exactly as described in the prompts for querying Large Language Models and provides all the listed files for the intermediate and final steps. Along with this reference implementation, PyTest scripts for all selected test case scenarios were prepared to facilitate evaluation and are available in the same repository \cite{Complex_benchmark}. 

\subsubsection{Technical verification} 
\label{sec:subsubsection}
The entire testing procedure described above assumes that the program generated by the Large Language Model will run successfully and generate all necessary files. Unfortunately, because of the scale of the chosen coding project, it may not always be the case. The LLMs may make minor mistakes that cause the application to fail during compilation. Such mistakes can include missing import statements or improper variable names that were changed between generating answers, but only in some places. Since the whole point of asking a model to solve a complex project was to bring the evaluation closer to real-life use of LLMs for coding tasks, disregarding the solution due to a few minor mistakes would contradict this idea. In such a case, a programmer working with the LLM would more likely review the code and manually fix the necessary fragments, or prompt the model until all mistakes are resolved. The second approach, unfortunately, undermines the experiment's reproducibility, as each model would need different prompts. Accordingly, the first approach was adopted as a preliminary step to verify the LLM-generated programs. Each code obtained from the models was copied one-to-one from the responses into the corresponding files, using the generated file names. After completing this setup, an attempt to run the program using the instructions provided by the LLM was made. All errors encountered during such attempts were documented and classified, and, to the best of the dedicated developer's knowledge, corrected. Since such manual corrections can introduce bias, the rules for making them were established. It was decided that each program should be fixed to a state in which there are no blocking syntax and runtime errors. The most challenging errors were those involving the logic of a solution, as they could differ significantly in complexity. Only mistakes that could be easily deduced from the error message itself were fixed. If the mistake required additional understanding or longer inference in the implementation, it was left out. Each error was assigned to one of the predefined categories, namely: syntax errors, compilation errors, logical errors, dependency errors, and others. While this technical verification step adds an extra layer of manual work to the presented procedure, it also provides additional information that can be used for model comparison.

\subsection{Developers study}
\label{sec:subsection}
After reviewing the basic aspects of LLM-generated code solutions, such as technical and logical correctness, it is time to investigate the second important area of code evaluation - the quality of the generated solution. This can include dimensions such as readability, maintainability, security, and usability. In many everyday scenarios of using LLMs for coding, a model takes the place of a developer, and a real human user sits in the seat of the ‘colleague’ who needs to understand, review, and accept the proposed solution. Therefore, it is good practice to keep code as clean and effective as possible, especially when working on an application that will be maintained over the long term or shared with other developers. Some aspects of code quality, such as well-used formatting or secure coding practices, can be verified using predefined checks. On the other hand, aspects such as easily understandable naming conventions or concisely written comments are equally important when working with the code, but may require a human eye to be properly evaluated. For the most valuable quality verification strategy, both approaches should be combined, as they complement one another.

\subsubsection{Automatic quality verification} 
\label{sec:subsubsection}
As mentioned, many of the improvements one can make to design a cleaner code are repeatable. This enabled the development of dedicated software solutions that search for predefined patterns in code and suggest ways to fix them.  One of the popular tools that offers such a solution is SonarQube software developed by SonarSource \cite{sonarsource_software_home}, which was selected for this methodology. It provides a built-in list of rules for many different languages, which can be simply selected for verification in the code. For example, at the time of writing this paper, there are 391 active rules defined for Python.
Each rule is classified by two dimensions: severity and clean code attribute. The first one describes the importance of the issue and the impact of breaking the rule on the software. The second one focuses on assigning one of the four main attributes of clean code, as defined by the company: consistency, intentionality, adaptability, and responsibility. Based on the number of found errors, their severity, and group placement, a summary is provided in three main categories. The categories are reliability, maintainability, and security, with ratings from A to E, where higher letters indicate better results \cite{sonarsource_software_quality}. 
The coverage of commonly known mistakes identified by automatic verification is a good starting point for discussing the quality of code generated by LLMs. However, because it is based on predefined rules, it cannot detect all areas for improvement, such as poorly defined method names, misleading comments, or incompatibility with the requirements, which are undetected by the tests. That is why a human eye is still needed in the code review process, as it can provide new, insightful information about the solution that was not detected in any of the previous evaluation steps.

\subsubsection{Human-based quality verification} 
\label{sec:subsubsection}
Usually, code reviews are conducted as free-text comments, which can focus on a small piece of code or the whole solution. This is a natural way to communicate between programmers and is often effective for verification by two or three developers. However, it may not be the best way to gather a range of opinions on several solutions and to compare them properly. To enable an objective comparison, each item should be rated using the same dimensions as those used for the previously mentioned evaluation aspects. Since this is not a standard way of conducting a code review, there is no standardised list of areas to evaluate when performing one. Nevertheless, because constant rating dimensions needed to be defined, such a list was designed specifically for this case. It is based on the author's experience performing and receiving code reviews in a corporate environment, with reference to some of the dimensions already defined by SonarQube and cross-validated against the dimensions used by Miah \& Zhu in their work on user-centric code evaluation \cite{UserCentricEval}. The resulting list comprises 14 aspects, divided into 3 categories, presented below. \\

Category "Structure of the code base":
\begin{itemize}
\item Names of files, classes, methods, and variables are reflecting on their purposes and make it easier to understand the code.
\item Formatting of the code makes it easily readable.
\item It is easy to understand what parameters are required by the functions
\item Type declarations and type hints are correctly used, allowing for easier code understanding.
\item Different functionalities are well separated from each other in different packages/classes/methods.
\item The code is not overengineered, and it does not include unnecessary operations
\end{itemize}

Category "Quality of comments and explanations":
\begin{itemize}
\item Comments are not too long to easily read, and they are not interfering with the code.
\item Comments are made in necessary places, where they have a positive impact on code understanding and readability.
\item Comments provide a good explanation of the code and allows for faster understanding.
\item Comments seems auto generated and not meaningful.
\item README file provides useful information about the code.
\end{itemize}

Category "Ease of usage":
\begin{itemize}
\item It would be easy to use functions defined in the code.
\item It would be easy to modify functions defined in the code.
\item It would be easy to test this code.
\end{itemize}
Each dimension focuses on one element, which can be analysed and rated independently. As can be seen, the metrics are defined as affirmative sentences to which one can either agree or disagree. After reviewing the code and reading the statements, the reviewers should indicate their level of agreement or disagreement with each statement on a 5-point Likert scale. The possible options are sorted from most negative to most positive: strongly disagree, disagree, neither agree nor disagree, agree, and strongly agree. This type of answer choice should ensure sufficient granularity without overwhelming reviewers and, at the same time, allow objective comparison of the quality of different programs in the eyes of real developers.

\subsubsection{Questionnaire setup} 
\label{sec:subsubsection}
After defining the metrics for human-based quality verification of LLM-generated code, a method for collecting the data must be designed. The methodology for gathering structured code reviews from developers should provide a way to easily share, access, and verify code samples, as well as to implement Likert-scale questions and collect participants' responses. Additionally, it should allow voluntary contributions from interested developers and ensure the anonymity of collected data. 

To simplify sharing LLM-generated code with evaluators, all programs were placed in a dedicated GitHub repository, with each generated solution on its own branch. To attract a wide range of participants to the study, the most effective approach is to offer the option to answer prepared questions online. The easiest approach would be to implement the defined metrics as part of the GitHub Pull Request process, which developers are accustomed to. Unfortunately, at the time of writing this paper, it does not seem possible. Because of this, other means of gathering user feedback needed to be chosen, and the decision fell on Google Forms, which allows users to create surveys for free and share them via a public link. The use of such a questionnaire-building tool also facilitates adherence to ethical research practices, ensuring informed consent, participant anonymity, voluntary participation, and data confidentiality.

At the beginning of the designed questionnaire, all necessary information about the project solved by the LLMs and the expected responses is provided. Such data are placed as a description of the first section in the questionnaire. Also, at the beginning, a disclaimer regarding the anonymity of answers and the scope of use is added. Afterwards, a brief section is added to collect information about the evaluator profile. The questions in this part relate to prior experience with LLMs and professional experience, which may influence the reviewer's responses. Since this section concerns participants' private information, all questions are optional.

This introductory part is followed by sections dedicated to answering Likert-scale questions. The number of such sections should be equal to the number of models one wants to evaluate and compare. Each section begins with a link to the PR containing the solution generated by the specified LLM, along with an overview of all provided files. Afterwards, the questions about the defined metrics are presented as a single-choice grid, with each statement allowing selection of only one option. Three such grids are created, one for each of the three categories: the structure of the code base, the quality of comments and explanations, and the ease of use. At the end of the section, an additional optional free-text field is provided that reads, \textit{If you want to leave any additional comments regarding the code, you can do it here}. This way, participants have a place to include all other comments that are important to them and do not fit predefined metrics. Results gathered in this field will not be used solely to compare the LLM's capabilities, but rather as an overview of developers' opinions of the provided code. At the end of the whole questionnaire, in the last field, the user is asked the following question: \textit{If you had to work on one of the codes generated for this task,  which code would you choose?}, where they must make a final decision on which model produced the best solution in their opinion.

\subsection{Methodology summary}
\label{sec:subsection}
To summarise, the created methodology is based on two main areas of code evaluation: correctness and quality. The flow starts with using predefined prompts to query the selected Large Language Model via API and saving the generated code. The program is then started, and during its run, the necessary output files are created. Based on the existence and content of those files, the correctness tests are performed, which allow to decide whether the program does what it was intended to do. After that, the code undergoes automated quality verification to determine whether it complies with a standardised set of quality rules. At the end, the code is shared with a group of volunteer developers, who evaluate it using a dedicated questionnaire which aims to reflect the code review process. This procedure should allow for measuring how useful the given program would be for developers to work with, including aspects such as its readability and ease of modification. The selected methodologies are complementary in scope and function, each addressing a specific area of code evaluation. This ensures a complete view of LLM's capabilities not only for solving a complex, multi-task assignment, but also for doing it in a way that is easy for developers to further work with. 

\section{Dedicated experiments}
To verify the feasibility of applying the proposed methodology in practice, a small set of dedicated experiments was conducted. The range of experiments included the evaluation of three general-purpose models: GPT-4.1, DeepSeek-V3-0324, and Claude Opus 4. Using 14 predefined prompts, the models were asked to solve the "Building the Tree of Life from Scratch” NIFTY assignment in Python. 

To fairly compare those coding projects, each model should use the same environment when generating the solution. To ensure these conditions, the experiment's author prepared a small script to request answers via available APIs. This automation ensured that not only were the prompts always the same, but also the models' parameters, such as the number of tokens and temperature, were fixed. Previous work has shown that lower temperatures are preferable for code generation tasks, especially when using a small number of samples \cite{humaneval}. Some sources, such as the DeepSeek documentation \cite{deepseek_temp}, even recommend setting the temperature to 0 for coding tasks. However, the project chosen for LLMs to solve is unique, and some of the designed prompts do not focus solely on code generation. Because of this, a slightly higher temperature was used to increase the likelihood of exploring the correct answers. For this work, the temperature is set to 0.2. and the maximum number of tokens the model can generate on the output is set to 4096. Due to this setup, the models' outputs are non-deterministic and may differ slightly when the same prompt is used multiple times. This aspect motivated the rationale that each model should be tested three times to properly measure its capabilities to solve the chosen assignment. For each run, all prepared prompts and LLM outputs were saved to a dedicated Markdown file. After the conversation ended, the file was manually examined, and code samples generated by the models were copied to separate files using the provided filenames. After all the code files were prepared, the solution was manually run by the author according to the instructions provided by the LLM in the README file. In cases of any errors, they were manually corrected, and the type of mistake was recorded. After that, each solution underwent quality verification performed using SonarQube.

Regarding the human-based part of the examined methodology, it was decided that asking developers to review three different codes is already challenging, and requesting even more work from potential participants may result in a lower response rate. As a result, only one code from each model was selected for the final review. The decision was based primarily on the correctness of the solution, with the version that passed the most test cases chosen. In case of equal values, the quality scores were also taken into account. The questionnaire was opened on 24 July 2025 when it was published on the private LinkedIn account of the paper's authors. Participants had over a month to complete the forms, as the survey closed on 28 August 2025. In the meantime, the post was reposted 4 times, and the additional post was submitted to the public group called “Python Developers Community (moderated)”, where it received over 50000 views, 32 reactions, and 1 repost. 

\subsection{Experiments results}
The results of the performed experiments are discussed in detail in the subsections below. One is dedicated to the correctness verification of programs generated by Large Language Models, and the other covers findings obtained from developers' survey results.
\begin{table*}
  \caption{Correctness results obtained by the LLMs}
  \label{tab:results}
  \begin{tabular}{cccccccccc}
    \toprule
    Metric & GPTv1 & GPTv2 & GPTv3 & Claudev1 & Claudev2 & Claudev3 & DeepSeekv1 & DeepSeekv2 & DeepSeekv3\\
    \midrule
    Runnable & Yes & No & Yes & No & No & No & No & No & No\\
    Syntax errors & 0 & 0 & 0 & 0 & 0 & 0 & 1 & 0 & 1\\
    Compilation errors & 0 & 0 & 0 & 3 & 1 & 0 & 3 & 4 & 3\\
    Logical errors & 0 & 1 & 0 & 2 & 1 & 1 & 1 & 0 & 0\\
    Dependency errors & 0 & 0 & 0 & 0 & 0 & 0 & 0 & 0 & 0\\
    Others errors & 0 & 0 & 0 & 0 & 0 & 1 & 0 & 0 & 0\\
    Security score & A & A & A & A & A & A & A & A & A\\
    Reliability score & A & A & A & A & A & A & A & A & A\\
    Maintainability score  & A & A & A & A & A & A & A & A & A\\
    Code smells & 3 & 6 & 5 & 11 & 4 & 9 & 1 & 2 & 6\\
    Debt ratio (\%) & 0.3 & 0.7 & 1.1 & 0.2 & 0.1 & 0.2 & 0 & 0.1 & 0.3\\
    Time to fix (min)  & 29 & 64 & 49 & 41 & 21 & 45 & 1 & 10 & 27\\
    Test cases passed & 4 & 4 & 3 & 2 & 0 & 3 & 3 & 2 & 5\\
  \bottomrule
\end{tabular}
\end{table*}

\subsubsection{Correctness results} 
\label{sec:subsubsection}
As shown in the Table~\ref{tab:results}, most of the obtained solutions could not run without intervention, except for two of the three outputs generated by GPT-4.1. However, the number of errors found was not too high, ranging from 1 to 5. Overall, the solutions with the fewest mistakes were generated by GPT-4.1, while those from DeepSeek required the most corrections. When it comes to the automatic quality verification results,  all codes got a score ‘A’ in security, reliability, and maintainability categories. The number of found code-smell cases was relatively low, with a maximum of 11 cases in one of the Claude solutions, which constituted 1.1\% of the codebase. All other solutions did not reach even 1\% of ‘smelly’ parts, but the worst in terms of time needed to fix those issues was one of the GPT-based codes, estimated at 64 minutes. Based on this, it can be concluded that, in terms of quality, the codes generated by the DeepSeek model are the best, as they have the fewest code-smell issues and the shortest estimated time to fix them, which does not exceed half an hour in any case. Most issues were found in Claude-based programs; however, GPT-based programs have the highest debt ratio and the longest time to fix issues. These results place GPT-4.1 as the LLM that generates the fewest high-quality solutions among the three selected models.

Many of the attempts failed the same tests, and none of the solutions could pass two of the prepared scenarios. The first one is a verification of the Needleman-Wunsch algorithm, where some models made a logical mistake by calculating scores for pairs that were created by taking the same species twice, for example, comparing a horse with itself. Such a pairing does not make sense from the point of view of the assignment, but the prompt used to generate that function contains the wording \textit{“for each pair of species calculate a Needleman-Wunsch similarity score”}, which could make it hard to deduce that pairs should be of different species. A similar situation is observed with the second test, none of the solutions passed, which is the one that compares created trees based on both node placement and distances between them. An observation made when examining files with those scores was that many of the values were negative, leading to the asssumption that branch lengths might be computed in the wrong direction. This confusion may be due to the fact that the prompt did not provide clear instructions on how the scores should be calculated, and only stated that the distances are expected. This could leave the LLM in a situation where it had to guess what to implement, and none of the models could make correct assumptions about this part. In addition to the two fully failed cases, LLMs also had difficulty building a correct tree, comparing it with the reference one using only node placement, and generating a working solution for the clustering task. In many cases, all possible species were assigned to a single cluster, regardless of the specified threshold. Only two cases, one from GPT and one from DeepSeek, have generated solutions that properly passed these two tests. Overall, the LLM that generated the best working solution in Python was DeepSeekV3-0324, which, on its third attempt, achieved a test pass score of 5 out of 7. A bit lower, but more stable results were provided by GPT-4.1, whose two solutions passed 4 test cases. Claude Opus 4 performed the worst in terms of correctness, with one attempt passing 0 tests and the remaining two passing only 2 and 3 tests, respectively.

Based on this part of the analysis, it can be concluded that the DeepSeek model comes out on top. It generated the best solution correctness-wise, as well as having the best results when it comes to the quality of the code.

\subsubsection{Questionnaire results} 
\label{sec:subsubsection}
As previously discussed, to avoid overwhelming participants with the amount of code to evaluate, only one solution from each model was selected for verification by developers. Based on the correctness evaluation results, for the GPT model, the code from attempt number 2 was chosen; for Claude, attempt number 1; and for DeepSeek, attempt number 3. Sadly, this modification and the high LinkedIn reach did not translate into a high response rate, as the questionnaire received only 10 responses. Because of this, the results from those experiments should be treated more as exploratory insights rather than statistically generalizable findings. Nevertheless, even this number of answers provides an interesting perspective on how people who work with code on a daily basis view different aspects of the prepared solutions. The raw, anonymous survey results are available to examine in a dedicated repository \cite{Survey_Results}.

From the introduction, which focused on collecting participants' metrics, it can be deduced that the respondents are mostly experienced IT employees working in large international companies with more than 1500 employees. All of them have at least small experience in using LLMs for code generation tasks, and in 30\% they are using it regularly. All but one person tried using GPT-4.1 in some way, but only 3 and 4 people used DeepSeek-V3 and Claude Opus 4, respectively. Although most people were familiar only with the GPT-4.1 model, it can surely be said that this model received the worst reviews in the survey. Based on the analysis of the Likert-scale questions about the GPT-4.1 code, the most negative responses were given to questions focused on the quality of comments, including placement, usefulness, and meaningfulness. Also, responses to questions about ease of modification and testing, naming conventions, and the separation of functionality were mostly negative or neutral. Additionally, this model received the most responses to the open-ended question, in which respondents commented on issues such as poor comments and a lack of documentation, a chaotic code structure, and hard-coded arguments instead of flexible configuration. Claude- and DeepSeek-based solutions were generally more positively perceived, having mostly agreed-upon answers. The Claude-generated code has received mostly neutral responses to questions about code over-engineering, comment placement, and comment meaningfulness. DeepSeek, on the other hand, received highly polarised responses regarding the code's over-engineering, with a majority of "neither agree nor disagree" responses on aspects of readability, ease of modification, and ease of testing. Both solutions had significantly fewer comments in the "additional comments" field, indicating better code structure and clearer comments. For DeepSeek, the better parameterisation of script parameters was also noted. This is a really important aspect of adding an open-ended question, as it allowed for identifying an area for improvement, which was not included in any previous evaluation step.
In the end, when asked about which code they would choose if they had to work with it, more users pointed to the Claude-generated one -  six people have chosen it, while only four decided on the DeepSeek-based program.

The survey results are in agreement with the correctness and quality results when it comes to solutions generated by GPT-4.1, placing it as the worst ones. However, while automated evaluations point to the DeepSeek model as the best one, the human-based survey results show more persuasion toward a Claude-originated solution. This may indicate that this program, even with a higher number of mistakes, was written in a more developer-friendly way.

\section{Discussion}
In the sections presented above, the detailed description of particular elements of the designed three-fold evaluation methodology for code generated by Large Language Models is provided. The development of this methodology and the experiments performed with it allow to formulate answers for all three research questions posed at the beginning of this paper.

The first question stated: \textit{How can we design an evaluation framework that assesses LLM-generated code in realistic, multi-component project contexts, rather than isolated function-level tasks?} For this problem, the proposed methodology assumes the use of Large Language Models to solve the assignment in a way most human users would: asking the model for additional pieces of the solution, prompt by prompt, and then running the entire solution at the end. After creating a solution, its correctness can be evaluated using a small set of unit tests that compare the expected final output and the saved intermediate results. 

The second research question addresses the problem of \textit{How to combine automated correctness testing and human quality assessment to comprehensively evaluate LLM-generated project implementations?} One way to do this is to ask programmers to evaluate the code generated by LLMs in the same way they often do during code reviews for colleagues. To standardise the procedure for gathering such code reviews, a web-based questionnaire with dedicated quality metrics measured on a Likert scale was designed. This way, it is possible to compare answers obtained while evaluating different LLMs. However, to not restrict participants in expressing their opinions and to keep the possibility of findings beyond the prepared metrics, an open-ended question was also included in the design. 

The third research question: \textit{To what extent automated correctness metrics and human quality assessments align and complement each other when evaluating LLM-generated projects?}, can be answered using the results of experiments conducted on three LLMs. Based on the correctness evaluation and questionnaire analysis described in the previous sections, it can be said that the two parts of the evaluation methodology complement each other. The benchmark results alone would point to the DeepSeek model as the best for solving the selected assignment, as it had the highest percentage of passed test cases. The technical evaluation underscores that only the GPT model generated a solution that runs without intervention. Ultimately, the static analysis also selected DeepSeek codes as those with the fewest mistakes and Claude as having the most code-smell points. However, the conclusion from the questionnaire part of the methodology is that participants mostly selected the Claude-based code as the one they would prefer to work with, and gave more positive responses about its ease of modification and testing. This may indicate that, from a human perspective, it may be better to work with code that is less correct but better structured and more readable. The experiment also showed the importance of giving participants the option to write their overall thoughts, as the open-ended comments highlighted that many aspects of the codes were not designed in a production-friendly way.

\section{Threats to validity}
Although the proposed methodology aims to bridge two important research gaps in the landscape of frameworks for evaluating LLM-based code generation, it is not without its own limitations. One of such limitations is the poor scalability of generating and running the coding project. Many parts of the work, such as copying code samples from LLM responses into relevant files or correcting runtime errors, were performed manually. Those steps were time-consuming and, in some cases, error-prone, since they depend on the level of focus of the person performing them. This may result in low interest in using this approach to verify the capabilities of other Large Language Models. However, an interesting variation of this study may be to use a dedicated tool, such as Microsoft Copilot, to automate this part of the experiment by writing the code generated by LLMs directly into the project files.  

The second area of limitation is related to the survey-based part of the methodology, as it shown to gather low interest from the developers. It may be beneficial to search for other platforms where volunteer developers can be found, or to evaluate only one model in a single survey, making it quicker to complete. Another aspect to consider is the representativeness of the participants, as for this experiment, they constituted a very specific group who must own a LinkedIn account, have programming knowledge, and be interested in completing surveys. Since the survey results strongly depend on participants' choices and the respondent groups in this study are quite narrow, it may be difficult to identify a corresponding group for other cases. However, an idea to encourage more participants to take such surveys may lie in popularising the methodology, educating about its importance, and promoting participation.

A final aspect that could further enrich the described study is the implementation of automation to analyse the collected questionnaire responses. Since the experiments described in the paper yielded only 10 responses, their overview and summary could be easily performed manually. However, if this study design were used at a larger scale, both in terms of the number of participants and the number of evaluated models, such manual work would no longer be feasible. Ideally, a metric that clearly addresses each dimension evaluated in the questionnaire and enables direct model comparison should be implemented. 

\section{Conclusion}
Complete evaluation of Large Language Models for code generation requires a comprehensive approach that captures both functional correctness and code quality. Existing methodologies often provide limited insight into real-world applicability and overlook the human element in the process.

This paper presented a three-fold evaluation methodology that addresses these limitations by combining project-level testing with automated quality overview and a structured developer survey. Additionally, a test of this methodology was conducted by using three Large Language Models to generate a Python-based solution for a complex computer science project.

The results of the performed experiments yielded several findings. It was demonstrated that project-level evaluation can be effectively implemented using carefully designed task descriptions and comprehensive test suites. Additionally, it was shown that using LLM-generated solutions as the basis for an online questionnaire and gathering responses on the quality of those programs can be used as a complete evaluation framework. Finally, it was shown that automated testing and developers' opinions complement each other, highlighting distinct areas for improvement.

The primary contribution of this work is a methodology applicable to current and future LLM evaluations. Additionally, each of the designed components, namely the prompts set and benchmark for correctness testing, as well as the prepared survey design, can serve as a separate aspect for evaluating the models. However, as proved by the experiments, the most comprehensive and broad results, which enable understanding of the capabilities and limitations of using LLMs for code generation, are obtained when applying all three parts of the methodology to the same solution.


\bibliographystyle{ACM-Reference-Format}
\bibliography{bibliography.bib}

@misc{austin2021MBPP,
      title={Program Synthesis with Large Language Models}, 
      author={Jacob Austin and Augustus Odena and Maxwell Nye and Maarten Bosma and Henryk Michalewski and David Dohan and Ellen Jiang and Carrie Cai and Michael Terry and Quoc Le and Charles Sutton},
      year={2021},
      eprint={2108.07732},
      archivePrefix={arXiv},
      primaryClass={cs.PL},
      url={https://arxiv.org/abs/2108.07732}, 
}

@misc{humaneval,
      title={Evaluating Large Language Models Trained on Code}, 
      author={Mark Chen and Jerry Tworek and Heewoo Jun and Qiming Yuan and Henrique Ponde de Oliveira Pinto and Jared Kaplan and Harri Edwards and Yuri Burda and Nicholas Joseph and Greg Brockman and Alex Ray and Raul Puri and Gretchen Krueger and Michael Petrov and Heidy Khlaaf and Girish Sastry and Pamela Mishkin and Brooke Chan and Scott Gray and Nick Ryder and Mikhail Pavlov and Alethea Power and Lukasz Kaiser and Mohammad Bavarian and Clemens Winter and Philippe Tillet and Felipe Petroski Such and Dave Cummings and Matthias Plappert and Fotios Chantzis and Elizabeth Barnes and Ariel Herbert-Voss and William Hebgen Guss and Alex Nichol and Alex Paino and Nikolas Tezak and Jie Tang and Igor Babuschkin and Suchir Balaji and Shantanu Jain and William Saunders and Christopher Hesse and Andrew N. Carr and Jan Leike and Josh Achiam and Vedant Misra and Evan Morikawa and Alec Radford and Matthew Knight and Miles Brundage and Mira Murati and Katie Mayer and Peter Welinder and Bob McGrew and Dario Amodei and Sam McCandlish and Ilya Sutskever and Wojciech Zaremba},
      year={2021},
      eprint={2107.03374},
      archivePrefix={arXiv},
      primaryClass={cs.LG},
      url={https://arxiv.org/abs/2107.03374}, 
}

@inproceedings{ClassEval,
author = {Du, Xueying and Liu, Mingwei and Wang, Kaixin and Wang, Hanlin and Liu, Junwei and Chen, Yixuan and Feng, Jiayi and Sha, Chaofeng and Peng, Xin and Lou, Yiling},
title = {Evaluating Large Language Models in Class-Level Code Generation},
year = {2024},
isbn = {9798400702174},
publisher = {Association for Computing Machinery},
address = {New York, NY, USA},
url = {https://doi.org/10.1145/3597503.3639219},
doi = {10.1145/3597503.3639219},
booktitle = {Proceedings of the IEEE/ACM 46th International Conference on Software Engineering},
articleno = {81},
numpages = {13},
location = {Lisbon, Portugal},
series = {ICSE '24}
}

@misc{humanevalplus,
      title={Is Your Code Generated by ChatGPT Really Correct? Rigorous Evaluation of Large Language Models for Code Generation}, 
      author={Jiawei Liu and Chunqiu Steven Xia and Yuyao Wang and Lingming Zhang},
      year={2023},
      eprint={2305.01210},
      archivePrefix={arXiv},
      primaryClass={cs.SE},
      url={https://arxiv.org/abs/2305.01210}, 
}

@misc{MBXP,
      title={Multi-lingual Evaluation of Code Generation Models}, 
      author={Ben Athiwaratkun and Sanjay Krishna Gouda and Zijian Wang and Xiaopeng Li and Yuchen Tian and Ming Tan and Wasi Uddin Ahmad and Shiqi Wang and Qing Sun and Mingyue Shang and Sujan Kumar Gonugondla and Hantian Ding and Varun Kumar and Nathan Fulton and Arash Farahani and Siddhartha Jain and Robert Giaquinto and Haifeng Qian and Murali Krishna Ramanathan and Ramesh Nallapati and Baishakhi Ray and Parminder Bhatia and Sudipta Sengupta and Dan Roth and Bing Xiang},
      year={2023},
      eprint={2210.14868},
      archivePrefix={arXiv},
      primaryClass={cs.LG},
      url={https://arxiv.org/abs/2210.14868}, 
}

@misc{HumanEval-X,
      title={CodeGeeX: A Pre-Trained Model for Code Generation with Multilingual Benchmarking on HumanEval-X}, 
      author={Qinkai Zheng and Xiao Xia and Xu Zou and Yuxiao Dong and Shan Wang and Yufei Xue and Zihan Wang and Lei Shen and Andi Wang and Yang Li and Teng Su and Zhilin Yang and Jie Tang},
      year={2024},
      eprint={2303.17568},
      archivePrefix={arXiv},
      primaryClass={cs.LG},
      url={https://arxiv.org/abs/2303.17568}, 
}

@inproceedings{CoderEval,
author = {Yu, Hao and Shen, Bo and Ran, Dezhi and Zhang, Jiaxin and Zhang, Qi and Ma, Yuchi and Liang, Guangtai and Li, Ying and Wang, Qianxiang and Xie, Tao},
title = {CoderEval: A Benchmark of Pragmatic Code Generation with Generative Pre-trained Models},
year = {2024},
isbn = {9798400702174},
publisher = {Association for Computing Machinery},
address = {New York, NY, USA},
url = {https://doi.org/10.1145/3597503.3623316},
doi = {10.1145/3597503.3623316},
booktitle = {Proceedings of the IEEE/ACM 46th International Conference on Software Engineering},
articleno = {37},
numpages = {12},
location = {Lisbon, Portugal},
series = {ICSE '24}
}

@misc{CONCODE,
      title={Mapping Language to Code in Programmatic Context}, 
      author={Srinivasan Iyer and Ioannis Konstas and Alvin Cheung and Luke Zettlemoyer},
      year={2018},
      eprint={1808.09588},
      archivePrefix={arXiv},
      primaryClass={cs.CL},
      url={https://arxiv.org/abs/1808.09588}, 
}

@misc{repobench,
      title={RepoBench: Benchmarking Repository-Level Code Auto-Completion Systems}, 
      author={Tianyang Liu and Canwen Xu and Julian McAuley},
      year={2023},
      eprint={2306.03091},
      archivePrefix={arXiv},
      primaryClass={cs.CL},
      url={https://arxiv.org/abs/2306.03091}, 
}

@misc{swebench,
      title={SWE-bench: Can Language Models Resolve Real-World GitHub Issues?}, 
      author={Carlos E. Jimenez and John Yang and Alexander Wettig and Shunyu Yao and Kexin Pei and Ofir Press and Karthik Narasimhan},
      year={2024},
      eprint={2310.06770},
      archivePrefix={arXiv},
      primaryClass={cs.CL},
      url={https://arxiv.org/abs/2310.06770}, 
}

@inproceedings{repoeval,
    title = "{R}epo{C}oder: Repository-Level Code Completion Through Iterative Retrieval and Generation",
    author = "Zhang, Fengji  and
      Chen, Bei  and
      Zhang, Yue  and
      Keung, Jacky  and
      Liu, Jin  and
      Zan, Daoguang  and
      Mao, Yi  and
      Lou, Jian-Guang  and
      Chen, Weizhu",
    editor = "Bouamor, Houda  and
      Pino, Juan  and
      Bali, Kalika",
    booktitle = "Proceedings of the 2023 Conference on Empirical Methods in Natural Language Processing",
    month = dec,
    year = "2023",
    address = "Singapore",
    publisher = "Association for Computational Linguistics",
    url = "https://aclanthology.org/2023.emnlp-main.151/",
    doi = "10.18653/v1/2023.emnlp-main.151",
    pages = "2471--2484"
}

@inproceedings{NIFTY_study,
author = {McDanel, Bradley and Novak, Ed},
title = {Designing LLM-Resistant Programming Assignments: Insights and Strategies for CS Educators},
year = {2025},
isbn = {9798400705311},
publisher = {Association for Computing Machinery},
address = {New York, NY, USA},
url = {https://doi.org/10.1145/3641554.3701872},
doi = {10.1145/3641554.3701872},
booktitle = {Proceedings of the 56th ACM Technical Symposium on Computer Science Education V. 1},
pages = {756–762},
numpages = {7},
location = {Pittsburgh, PA, USA},
series = {SIGCSETS 2025}
}

@INPROCEEDINGS{UserCentricEval,
  author={Miah, Tanha and Zhu, Hong},
  booktitle={2024 IEEE International Conference on Artificial Intelligence Testing (AITest)}, 
  title={User Centric Evaluation of Code Generation Tools (Invited Paper)}, 
  year={2024},
  volume={},
  number={},
  pages={109-119},
  doi={10.1109/AITest62860.2024.00022}}

@article{WWang2024,
author = {Wang, Wei and Ning, Huilong and Zhang, Gaowei and Liu, Libo and Wang, Yi},
title = {Rocks Coding, Not Development: A Human-Centric, Experimental Evaluation of LLM-Supported SE Tasks},
year = {2024},
issue_date = {July 2024},
publisher = {Association for Computing Machinery},
address = {New York, NY, USA},
volume = {1},
number = {FSE},
url = {https://doi.org/10.1145/3643758},
doi = {10.1145/3643758},
journal = {Proc. ACM Softw. Eng.},
month = jul,
articleno = {32},
numpages = {23}
}

@misc{judgingllmasajudge,
      title={Judging LLM-as-a-Judge with MT-Bench and Chatbot Arena}, 
      author={Lianmin Zheng and Wei-Lin Chiang and Ying Sheng and Siyuan Zhuang and Zhanghao Wu and Yonghao Zhuang and Zi Lin and Zhuohan Li and Dacheng Li and Eric P. Xing and Hao Zhang and Joseph E. Gonzalez and Ion Stoica},
      year={2023},
      eprint={2306.05685},
      archivePrefix={arXiv},
      primaryClass={cs.CL},
      url={https://arxiv.org/abs/2306.05685}, 
}

@misc{farchi2024,
      title={Automatic Generation of Benchmarks and Reliable LLM Judgment for Code Tasks}, 
      author={Eitan Farchi and Shmulik Froimovich and Rami Katan and Orna Raz},
      year={2024},
      eprint={2410.21071},
      archivePrefix={arXiv},
      primaryClass={cs.SE},
      url={https://arxiv.org/abs/2410.21071}, 
}

@article{Can_LLMs_Replace_Human_Evaluators,
author = {Wang, Ruiqi and Guo, Jiyu and Gao, Cuiyun and Fan, Guodong and Chong, Chun Yong and Xia, Xin},
title = {Can LLMs Replace Human Evaluators? An Empirical Study of LLM-as-a-Judge in Software Engineering},
year = {2025},
issue_date = {July 2025},
publisher = {Association for Computing Machinery},
address = {New York, NY, USA},
volume = {2},
number = {ISSTA},
url = {https://doi.org/10.1145/3728963},
doi = {10.1145/3728963},
journal = {Proc. ACM Softw. Eng.},
month = jun,
articleno = {ISSTA086},
numpages = {23}
}

@online{NIFTY_webpage,
	title = {Building The Tree of Life from Scratch},
	url = {http://nifty.stanford.edu/2025/tralie-phylogenetic-trees/},
	author = {Tralie, Christopher},
	urldate = {2025-08-25},
	date = {2025},
}

@dataset{Survey_Results,
  author       = {Szych, Joanna},
  title        = {Evaluating LLM-Generated Code: Developer Study},
  month        = feb,
  year         = 2026,
  publisher    = {Zenodo},
  version      = {pre-review},
  doi          = {10.5281/zenodo.18806359},
  url          = {https://doi.org/10.5281/zenodo.18806359},
}

@software{Complex_benchmark,
author = {Szych, Joanna},
month = feb,
title = {Evaluating LLM-Generated Code: Benchmarking on complex assignment},
version = {1.0.0},
year = {2026},
url = {https://github.com/AsiaSzych/Tree_of_Life/}
}

@online{sonarsource_software_quality,
	title = {Software qualities {\textbar} {SonarQube} Cloud Documentation},
	url = {https://docs.sonarsource.com/sonarqube-cloud/digging-deeper/software-qualities/},
	author = {{SonarSource}},
	urldate = {2025-08-27},
	date = {2025},
	langid = {english},
}

@online{sonarsource_software_home,
	title = {Homepage | SonarQube Cloud | Sonar Documentation},
	url = {https://docs.sonarsource.com/sonarqube-cloud},
	author = {{SonarSource}},
	urldate = {2025-08-27},
	date = {2025},
	langid = {english},
}

@online{deepseek_temp,
	title = {The Temperature Parameter {\textbar} {DeepSeek} {API} Docs},
	url = {https://api-docs.deepseek.com/quick_start/parameter_settings},
    author = {{DeepSeek}},
	urldate = {2025-09-09},
    date = {2025},
	langid = {english},
}

\appendix

\end{document}